\documentclass[runningheads]{llncs}

\usepackage{accv}

\usepackage{accvabbrv}

\usepackage{graphicx}
\usepackage{booktabs}
\usepackage{amsmath}
\usepackage{tabularx}
\usepackage{array}
\usepackage{makecell}
\usepackage{multirow}
\usepackage{colortbl}   
\usepackage{enumitem}

\usepackage[accsupp]{axessibility}  

\newcommand{\boldemph}[1]{\textbf{\textit{#1}}}

\usepackage{hyperref}

\usepackage{orcidlink}

\AtEndPreamble{%
  \crefname{figure}{Fig.}{Fig.}%
  \Crefname{figure}{Figure}{Figures}%
}

\begin{document}

\title{Lightweight and Resource-Efficient Perception for Robotic Guide Dogs}


\author{
\mbox{Jinse Kwon\inst{1}\orcidlink{0000-0003-3091-9926}} \and
\mbox{Yoojin Lim\inst{1}\orcidlink{0000-0002-8107-2553}} \and 
\mbox{Choonghan Lee\inst{1}\orcidlink{0000-0003-4883-6707}} \and
\mbox{Yongseung Yu\inst{1}\orcidlink{0000-0001-9838-4310}}  \and
\mbox{Yongin Kwon\inst{2}\orcidlink{0000-0003-2973-246X}}  \and
\mbox{Jemin Lee\inst{3,}$^\dagger$\orcidlink{0000-0002-9332-3508}}
}

\authorrunning{J. Kwon et al.}

\institute{Electronics and Telecommunications Research Institute (ETRI), Korea \\
\email{\{kwonse, yoojin.lim, choonghan.lee, ys.yu\}@etri.re.kr}
\and 
School of Computer Science and Engineering, Pusan National University, Korea \\
\email{yongin@pusan.ac.kr}
\and
Division of Electronics and Information Engineering, Jeonbuk National University, Korea\\
\email{jemin.lee@jbnu.ac.kr}
}

\maketitle

\begingroup

\renewcommand{\thefootnote}{$\dagger$}

\footnotetext{Corresponding author}

\endgroup

\renewcommand{\thefootnote}{\fnsymbol{footnote}}
\footnotetext[1]{Code available at: \url{https://github.com/JinseKwon/guidedog-perception}}


\begin{abstract}
Robotic guide dogs should understand their surroundings, objects, and potential risks.
Prior research has focused on raw sensor data from cameras and 2D or 3D LiDAR, which precisely measure distance points rather than provide a semantic understanding of the scene.
While these physical measurements are effective for robot-centric collision avoidance and robot safety, they are not suitable for human-centric guidance.
The system should recognize the type and relevance of obstacles and explain them, clearly and actionably, in terms of their spatial relation to the user.
We present complete on-device perception modules that fuse a 360 camera and a 2D LiDAR for reliable collision avoidance, with moving-object detection and tracking for human-centric guidance.
Finally, in walking-impossible situations, a vision--language model delivers pathway explanations as a safety mechanism to reduce user anxiety.
In experiments, verification of fused 360 camera--LiDAR depth shows reliable near-range perception but inherent mid-range bias, while the system as a whole sustained real-time performance under 55 W.
On the real-world egocentric GuideDogQA benchmark, our system achieved 83.8\% accuracy, compared with 67.1\% for GPT-4o.
These results demonstrate that practical human-centric guidance with real-time on-device inference is feasible even on quadrupeds.

\end{abstract}

\section{Introduction}
\label{sec:intro}

Mobility for blind and low-vision people hinges on reliable perception and safe, efficient locomotion.
Guide dogs give people with visual impairments greater independence and mobility, but access is severely limited.
Training requires years of effort and tens of thousands of dollars, while guide dogs can only work for a short portion of their lives.
Because of these constraints, only about 2--3\% of those who could benefit are ever matched with a guide dog~\cite{kim2023_quadruped_guide,hwang2024_chi,cohav2025_looks_matter}.

These limitations have motivated efforts to develop robotic guide dogs as an alternative.
Unlike biological dogs, robotic guides should not only ensure their own locomotion safety but also detect hazards relevant to human users~\cite{kim2023_quadruped_guide}.
Potential dangers include overhanging obstacles above the head, surface irregularities or damages near the ground, as well as dynamic obstacles approaching from the sides or even the rear.
A robotic guide therefore requires omnidirectional sensing, accurate distance measurement, and stable performance, all under strict limits on size, weight, power, and cost.

\begin{figure}[t]
  \centering
  \begin{subfigure}{0.43\linewidth}
    \includegraphics[width=\linewidth]{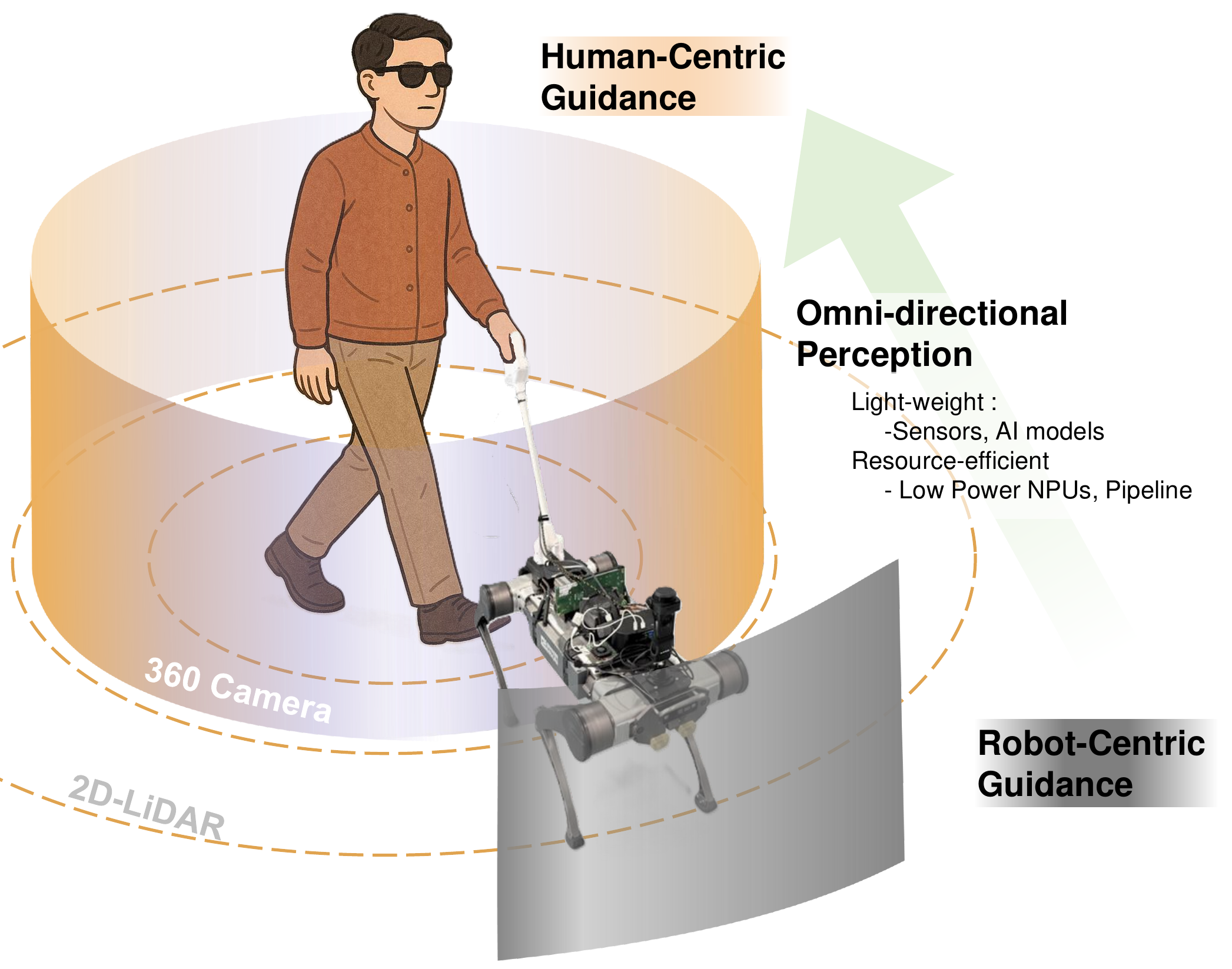}
    \caption{}
    \label{fig:fig1}
  \end{subfigure}
  \hfill
  \begin{subfigure}{0.43\linewidth}
    \includegraphics[width=\linewidth]{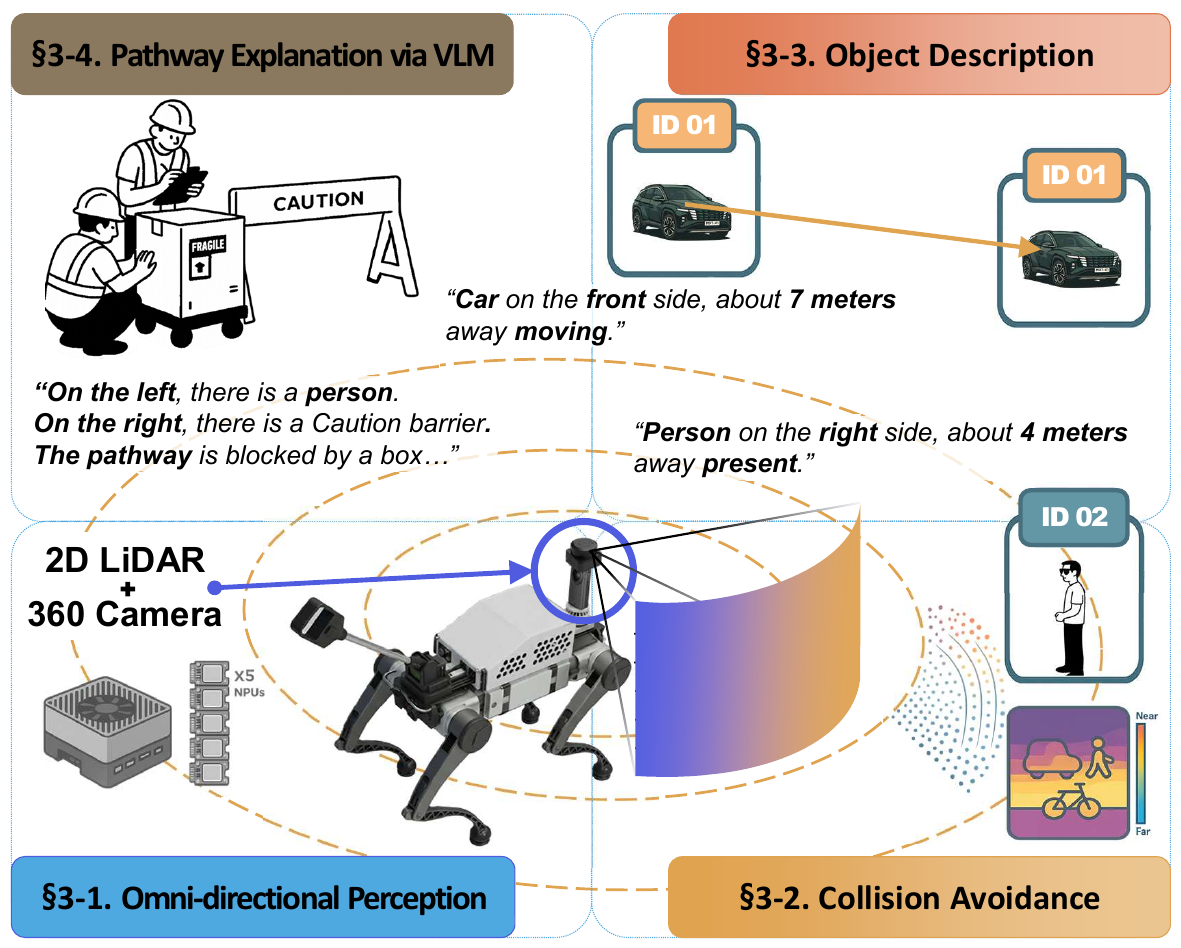}
    \caption{}
    \label{fig:pipeline_overview}
  \end{subfigure}
  \caption{(a)~A 360 camera and a 2D LiDAR fused with on-device AI and
  multi-NPU acceleration, moving from robot-centric toward human-centric guidance. (b)~Overview of the proposed system, integrating perception, obstacle understanding, and VLM-based guidance.}
  \label{fig:teaser}
\end{figure}

We develop omnidirectional perception modules for compact quadruped guide dogs and detail their algorithmic design and on-device performance. \cref{fig:teaser} summarizes the proposed platform and overall pipeline.

The handler expresses high-level intent, while the robot performs local path following and collision avoidance. Our system provides the omnidirectional geometry and human-centric scene understanding required for this division of labor.

Our contributions are:
\begin{itemize}[label=\textbullet]
  \item \textbf{On-device omnidirectional metric depth:}
   We propose a cube-map based omnidirectional depth pipeline that anchors per-face monocular depth with LiDAR-guided affine correction.

  \item \textbf{Human-centric perception and guidance:}
   We integrate moving-object perception and event-driven VLM--TTS explanation for human-centric guidance, achieving 83.8\% accuracy on GuideDogQA relative-depth comparison compared with 67.1\% for GPT-4o.

  \item \textbf{Real-time deployable system:}
  We deploy the complete stack on a Jetson AGX Orin with multiple Hailo-8 NPUs, sustaining real-time depth estimation at 25.3 FPS per cube-map face and object detection at 11.4 FPS per cube-map face within a 55 W power budget.
\end{itemize}

\section{Related Work}
\label{sec:related}

\noindent\textbf{Robotic guide dogs.}
Prior work spans safety formalization for quadruped guides~\cite{kim2023_quadruped_guide}, leash-guided physical human--robot interaction~\cite{xiao2021_leash_guided}, studies of how handlers and guide dogs cooperate and co-construct perception~\cite{hwang2024_chi,due2023_mobilities}, and the functional and aesthetic preferences of blind and low-vision (BLV) users~\cite{cohav2025_looks_matter}.
On the perception side, Hwang et al.~\cite{hwang2023_system_icra} built a camera-only system whose inference ran on a tethered laptop GPU. 
Even low-resolution segmentation was difficult to sustain on device. 
Aoki et al.~\cite{aoki2023_guidedog_3dlidar} combined a 3D LiDAR with a stereo depth camera, achieving accurate 3D geometry. 
However, the system is heavy, power hungry, and costs nearly ten thousand dollars. 
Moreover, it still provides only distance measurements rather than the semantic descriptions required for human-centric guidance of BLV users.
We instead pair a consumer 360 camera with a low-cost 2D LiDAR within a small quadruped's size--weight--power--cost budget, keeping every perception module on-device.

\noindent\textbf{Lightweight depth estimation and 360 fusion.}
Monocular depth estimation has evolved toward diverse accuracy--efficiency trade-offs. 
Representative methods include FastDepth~\cite{fastdepth} for embedded real-time applications, SC-DepthV3~\cite{scdepthv3} for self-supervised depth estimation in dynamic scenes, Depth-Anything-V2 (DA-V2)~\cite{depth-anyv2} for large-scale generalization, and the recently proposed Depth-Anything-V3~\cite{depth-anyv3}, which further improves zero-shot depth estimation.
However, these methods estimate only relative or scale-ambiguous depth, making their predictions insufficient for direct use in real-world navigation.
We adopt SC-DepthV3 for its native Hailo-8~\cite{hailo8} acceleration and recover metric scale using a per-face affine fit against the corresponding 2D LiDAR scan.
For 360 imagery, OmniFusion~\cite{omnifusion} removes projection distortion using tangent-image patches. 
However, it requires many overlapping inferences per frame. 
Instead, we retain the more efficient four-face cube-map decomposition, whose faces naturally align with the LiDAR's angular sectors. 
Residual boundary distortion is mitigated through overlapping views and seam-aware blending.

\noindent\textbf{VLMs for BLV spatial assistance.}
A consistent body of evidence shows that current VLMs have limited spatial reasoning capabilities. 
This limitation has been reported across embodied question answering~\cite{openeqa}, 3D language interfaces~\cite{pointllm,3dllm}, spatially fine-tuned VLMs~\cite{spatialvlm}, and even frontier models~\cite{gpt5si}.
We observed similar failures when prompting VLMs with structured spatial metadata. 
Therefore, our moving-object descriptions are generated geometrically, while language models are reserved for event-driven pathway explanation following the space-aware instruction tuning of SA-VLM~\cite{savlm}.
GuideDog~\cite{guidedog} is the closest work to our evaluation. 
It introduces a large egocentric BLV guidance dataset together with the GuideDogQA benchmark, on which we compare our deployed on-device perception modules against substantially larger VLMs.

\section{Omnidirectional Perception Modules}
No single affordable sensor meets our requirements: a monocular camera is rich in semantics but weak in depth and field of view, a 360 camera needs distortion handling, and a 2D LiDAR ranges accurately at low power but only in one plane.
We therefore combine a 360 camera with a 2D LiDAR and perform online fusion.

\begin{figure}[t]
  \centering
  \includegraphics[width=0.65\linewidth]{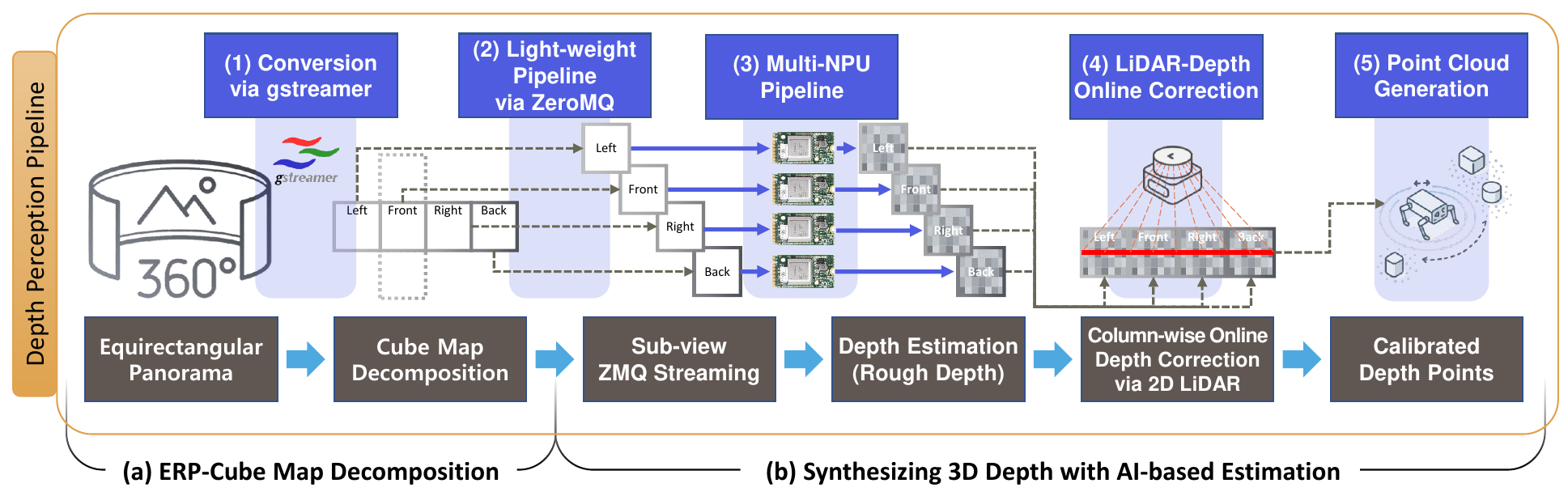}
  \caption{Omnidirectional depth pipeline: (1) gstreamer-based cube-map decomposition, (2) ZMQ streaming to parallel NPUs, (3) SC-DepthV3 depth estimation, (4) LiDAR-based affine correction, and (5) 3D point generation for collision avoidance.}
  \label{fig:coli-avoid}
\end{figure}

\subsection{Equirectangular to Cube map decomposition}
\label{subsec:erp}
We capture an equirectangular panorama (ERP) and forward it to a gstreamer pipeline, then decompose each frame into four perspective faces, namely front, left, back, and right, that together cover the entire horizontal view, as illustrated in \cref{fig:coli-avoid}.
We explicitly discard the top and bottom faces and restrict each face to a vertical field of view (VFOV) of $90^\circ$, which is sufficient for human-level clearance checks on a guide-dog platform and saves compute.
To accelerate processing, we precompute and cache mapping grids per resolution, ensuring linear runtime complexity in the number of pixels. When integrated into a gstreamer pipeline, the remapping further benefits from hardware decoder support, resulting in much faster performance on embedded hardware.
The faces are then published over ZeroMQ (ZMQ)~\cite{zmq} as independent streams, front, back, left, and right, so that downstream modules such as detection, depth, and tracking can subscribe selectively and run in parallel on separate NPUs.

\subsection{Synthesizing 3D Depth with AI-based Estimation}
\label{subsec:depth-fusion}
The cube map decomposition gives the robot omnidirectional vision, but RGB alone is insufficient for reliable clearance.
We therefore attach monocular depth estimation to each face to obtain a dense depth field for near-field guidance.
Among DA-V2, SC-DepthV3~\cite{scdepthv3}, and FastDepth~\cite{fastdepth}, we adopt SC-DepthV3 as the default estimator, since it offers a practical balance of accuracy, efficiency, and deployability with native acceleration on Hailo-8~\cite{hailo8}.
\cref{fig:coli-avoid} shows the complete pipeline for generating 3D point information by combining a 360 camera with a 2D LiDAR, which we describe step-by-step below.

\boldemph{(1) Per-face affine depth correction.}
For face columns $u=0,\dots,W-1$ we know the azimuth by construction,
\begin{equation}
\theta_u \;=\; \theta_{\text{start}} + \frac{u}{W-1}\,(\theta_{\text{end}}-\theta_{\text{start}}),
\label{eq:theta}
\end{equation}
and we subscribe to the matching LiDAR sector, where the 2D LiDAR measurements are also broadcast as four directional ZMQ streams by our module.
Let $x_u$ be the monocular depth on the mid-row, i.e., the horizontal center line of the estimated depth map, and $\tilde r_u$ be the LiDAR range interpolated at $\theta_u$.
A tiny, angle-dependent affine fit stabilizes scale,
\begin{equation}
\tilde r_u \;\approx\; a(\theta_u)\,x_u + b(\theta_u),
\label{eq:affine}
\end{equation}
with $a(\cdot),b(\cdot)$ represented by a few sinusoidal bases (constant, $\sin\theta$, $\cos\theta$).
Only six coefficients are solved per face to align the monocular depth with the 2D LiDAR measurements, which keeps the computation lightweight enough for real-time processing.

We specify \cref{eq:affine} concretely by parameterizing the angle-dependent gains with just three bases per term:
\begin{equation}
\begin{split}
a(\theta)=\beta_0+\beta_1\sin\theta+\beta_2\cos\theta,\\
b(\theta)=\beta_3+\beta_4\sin\theta+\beta_5\cos\theta.
\end{split}
\label{eq:abparam}
\end{equation}
Substituting \cref{eq:abparam} into \cref{eq:affine} turns each valid mid-row column $u$ into one linear observation
$\tilde r_u\approx\boldsymbol\phi_u^\top\boldsymbol\beta$ with
$\boldsymbol\phi_u=[x_u,x_u\sin\theta_u,x_u\cos\theta_u,1,\sin\theta_u,\cos\theta_u]^\top$,
where $\theta_u$ comes from \cref{eq:theta} and $x_u$ is the monocular depth at the mid row.
Given $M$ valid columns, we estimate the coefficients by ridge regression,
\begin{equation}
\hat{\boldsymbol\beta}
= \arg\min_{\boldsymbol\beta\in\mathbb{R}^6}
\sum_{u=1}^M \big(\boldsymbol\phi_u^\top \boldsymbol\beta - \tilde r_u\big)^2
+ \lambda \|\boldsymbol\beta\|^2.
\label{eq:ridge}
\end{equation}

Rather than materializing the $M\times6$ design matrix, we accumulate the sufficient statistics
$G=\sum_u\boldsymbol\phi_u\boldsymbol\phi_u^\top\in\mathbb{R}^{6\times6}$ and
$\mathbf h=\sum_u\boldsymbol\phi_u\tilde r_u$, and solve the resulting tiny normal
equations $(G+\lambda I)\,\hat{\boldsymbol\beta}=\mathbf h$, which takes only
microseconds per face and constant memory.

To stabilize frame-to-frame behavior, we smooth coefficients with an exponential moving average and gate updates when data are insufficient or residuals exceed a threshold.
Precomputing $\{\sin\theta_u,\cos\theta_u\}$ per face and unwrapping angles ensures robust interpolation.
This compact six-parameter controller grounds monocular depth to a real-world scale and applies the affine correction across the entire face, rescaling each pixel as $a(\theta_u)x_{u,v}+b(\theta_u)$.
In this way all rows share a consistent distance scale, yielding depth maps that are globally aligned yet lightweight enough for on-device use and accurate enough for guide-dog safety.

\boldemph{(2) Overlapping faces and seam smoothing.}
Although the per-face correction stabilizes the scale within each face, the depth values near face boundaries remain noticeably unstable.
To mitigate this effect, we construct four additional overlapping views, each formed by concatenating two adjacent faces with 50\% horizontal overlap, and apply the same regression procedure on these overlaps so that the affine correction is solved with shared support across neighboring faces.
The final coefficients are obtained by blending the per-face and overlapping estimates, yielding smoother transitions across seams.

After distance locking, we further smooth the seams by interpolating across a boundary of width $w$ pixels.
Here we parameterize the horizontal position within this seam by a normalized coordinate $t \in [0,1]$, rather than by angle, and apply a cosine ramp
\begin{equation}
\rho(t)=\tfrac{1}{2}\big(1-\cos(\pi t)\big).
\label{eq:cosine}
\end{equation}
The weight $\rho(t)$ smoothly increases from $0$ to $1$, and the fused depth strip is
\[
D_{\text{fused}}=\rho(t)\,D_{\text{ov}}+(1-\rho(t))\,D_{\text{base}},
\]
where $D_{\text{base}}$ is the per-face estimate and $D_{\text{ov}}$ is the overlapping estimate from the adjacent face.
The left seam uses the overlap with the previous face, the right seam uses the next face, while the center keeps the base result to avoid unnecessary computation.

\begin{figure}[t]
  \centering
  \includegraphics[width=0.65\linewidth]{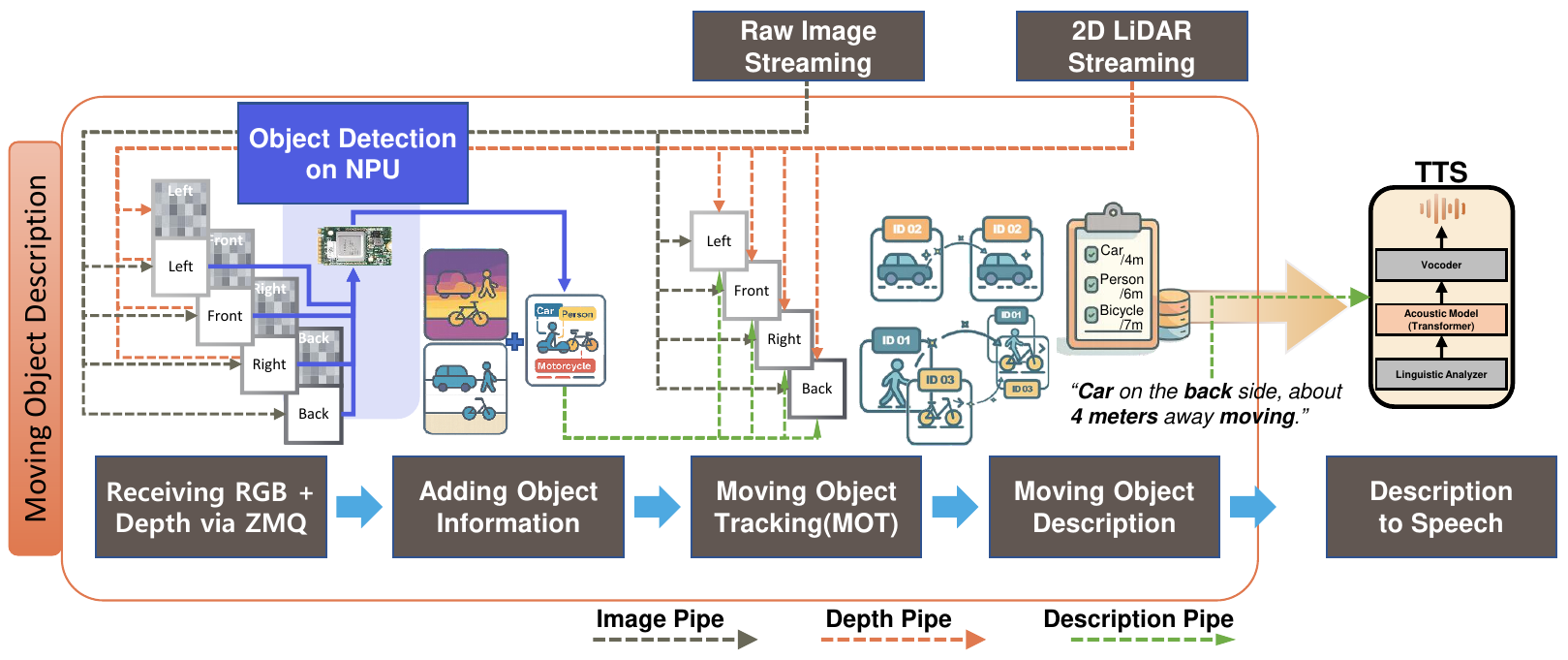}
  \caption{Moving-object description: four faces stream RGB--depth into NPU detection (YOLOv8m) and lightweight tracking for object kinematics; per-face tracks are aggregated with safety-based prioritization into one spoken description per cycle.}
  \label{fig:mod}
\end{figure}

\subsection{Moving Object Description}
\label{subsec:mo-desc}

Given four synchronized faces each provides an RGB frame
$\mathbf{I}\in\mathbb{R}^{480\times 480\times 3}$ and a depth map
$D\in\mathbb{R}^{256\times 320}$ where every pixel carries an estimated distance in millimeters.
We attach a lightweight detection and tracking head that transforms raw pixels into human-meaningful descriptions of the scene, as illustrated in \cref{fig:mod}.
It identifies the class of each object, estimates its position in terms of azimuth, elevation, and range, and characterizes its motion through radial speed and angular rates.
The module executes independently of the depth fusion stage and then aggregates the results across all four faces to generate short spoken summaries, which are delivered to the visually impaired user.

\boldemph{(1) Inputs and calibration.}
We crop a centered ROI in the RGB frame to match the depth aspect ratio (height $384$ in our implementation), then map RGB pixel coordinates $(x,y)$ to depth indices with fixed scales $s_x=\tfrac{320}{480}$, $s_y=\tfrac{256}{384}$ and a vertical offset $y_0$, i.e., $(\hat{x},\hat{y})=(\mathrm{round}(s_x x),\,\mathrm{round}(s_y(y-y_0)))$.
Azimuth and elevation are then computed from the horizontal and vertical fields of view (HFOV, VFOV):
\begin{equation}
\begin{split}
\theta_h(x) &= \Big(\tfrac{x+\tfrac12}{W}-\tfrac12\Big)\,\mathrm{HFOV}, \\
\theta_v(y) &= -\Big(\tfrac{(y-y_0)+\tfrac12}{H_{\mathrm{roi}}}-\tfrac12\Big)\,\mathrm{VFOV}.
\end{split}
\end{equation}
where $W=480$ and $H_{\mathrm{roi}}=384$. (The negative sign follows the standard image-downwards $y$ axis.)
Although HFOV and VFOV are in principle free parameters, we did not retrain the depth estimation backbone and simply used the pretrained model at $320\times256$, which fixes HFOV at $90^\circ$ and limits the VFOV of the depth ROI to $72^\circ$.

\boldemph{(2) Per-face detection.}
Each face executes a compact YOLO-family detector compiled for Hailo-8~\cite{hailo8}.
While any YOLO variant can in principle be supported, we adopt YOLOv8m~\cite{yolov8} as a practical trade-off between accuracy and efficiency on our hardware.
To avoid wasting resources on indiscriminate tracking, we restrict the set of classes to those relevant for mobility (person, bicycle, car, bus, truck, dog, cat, etc.), assuming typical pedestrian environments and discarding very small or highly specific classes that do not obstruct walking. For the same reason, we apply a confidence threshold of $0.8$ to suppress marginal detections.
For a detection $k$ with center $(x_k,y_k)$ we then read the aligned depth $d_k = D(\hat{x}_k,\hat{y}_k)$.

\boldemph{(3) Lightweight tracking and kinematics.}
We maintain per-class tracks with greedy nearest-neighbor association (pixel distance $<60$\,px, class must match) and a small miss counter to drop stale tracks.
For each track $i$ updated at time $t$ with depth $d_i(t)$ and bearings $(\theta_h(t),\theta_v(t))$, we estimate radial velocity $v_{d,i}$ and angular velocities $\omega_{h,i}$, $\omega_{v,i}$ using finite differences followed by exponential smoothing with factor $\gamma\!\in\![0,1)$ to reduce noise.
A simple rule flags objects as ``moving'' if $|v_{d,i}| \ge \tau_{\text{move}}$ (default $0.10$\,m/s).

\boldemph{(4) From four faces to a single description.}
Every $T_{\text{sum}}$ seconds (default $1$\,s) we aggregate the most recent tracks from all faces, following the pipeline shown in \cref{fig:mod}.
In our current implementation, we adopt a simple priority scheme that emphasizes the most critical cues for mobility: we describe only one sentence per cycle, giving precedence to the closest object, objects in the front direction, and vehicle-related classes such as cars and buses.
This prioritization ensures that the spoken output remains concise and focused on safety-critical elements.

We initially evaluated LLM and VLM based free-form descriptions of moving objects. However, they did not reliably identify moving or high-risk objects, even when provided with structured JSON spatial metadata. We therefore retain the deterministic priority scheme described above.

\begin{figure}[t]
  \centering
  \includegraphics[width=0.55\linewidth]{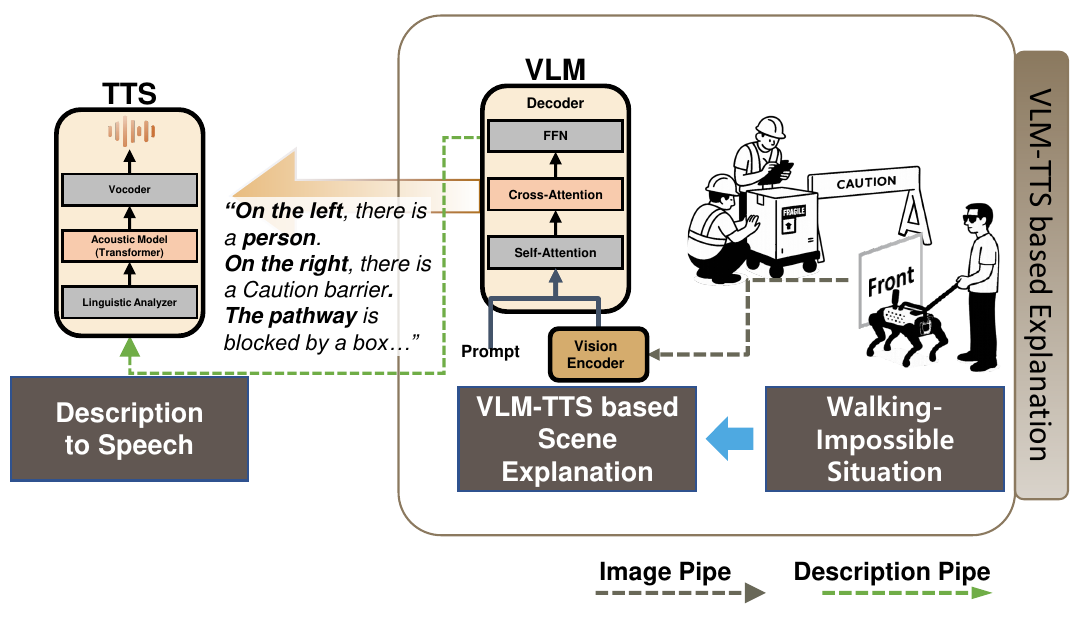}
  \caption{Event-driven pathway explanation: on a blocked path, a VLM turns the front view into a space-aware description of obstacles and options, streamed sentence-by-sentence to TTS to explain why walking stopped.}
  \label{fig:vlm-tts}
\end{figure}

\newcommand{\NA}{\textemdash}
\newcommand{\softhline}{
  \arrayrulecolor{black!15}\specialrule{.1pt}{0pt}{0pt}\arrayrulecolor{black}
}

\subsection{VLM-TTS Based Explanation}
\label{subsec:vlm-tts}

When the collision avoidance stack declares \texttt{walking\_impossible}, the robot first stops and generates a spoken pathway explanation. 
Following the space-aware guidance design~\cite{savlm}, the VLM describes relevant obstacles, their locations, and immediate options, explaining why the robot stopped and whether further walking is possible.

As illustrated in \cref{fig:vlm-tts}, the VLM analyzes only the front image and generates a spatially aware description of the forward scene, including left, right, and pathway context. The output is immediately delivered through Text-to-Speech (TTS).

Perception, VLM, and TTS are decoupled over a lightweight ZMQ message bus. Generated sentences are streamed to TTS to reduce first-utterance latency, while omnidirectional depth perception continues monitoring the surroundings. Once the path becomes clear, the explanation mode terminates and normal guidance resumes.

\begin{figure}[t]
  \centering
  \includegraphics[width=0.5\linewidth]{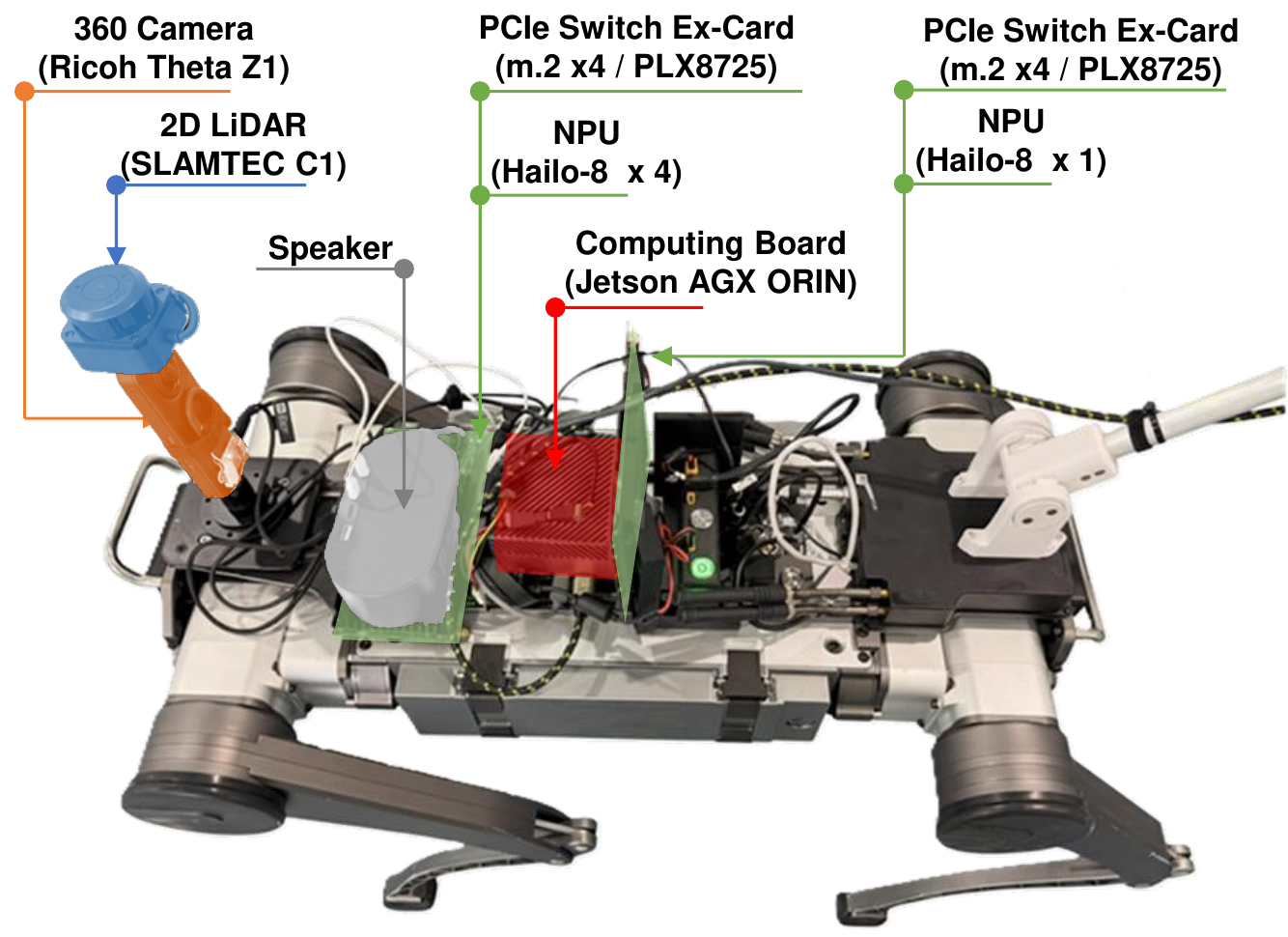}
  \caption{Hardware setup: Ricoh THETA Z1 360 camera, SLAMTEC C1 2D LiDAR, Jetson AGX Orin, and five Hailo-8 NPUs.}
  \label{fig:setup}
\end{figure}

\section{Experiments}

\subsection{On-device Platform and Configuration}
\label{subsec:exp-setup}

Our hardware configuration is illustrated in \cref{fig:setup}: an NVIDIA Jetson AGX Orin~\cite{orin}, a Ricoh THETA Z1 360 camera~\cite{thetaz1} (USB UVC), and a SLAMTEC C1 2D LiDAR~\cite{slamc1}.
The camera outputs equirectangular frames, which we split into four cube map faces: front, left, back, and right.
Each face is cropped to a $90^\circ$ vertical field of view, producing $480{\times}480$ BGR tiles for detection, along with $320{\times}256$ fused depth maps in millimeters.
The LiDAR operates at $\geq$10 Hz. 
Its scan is divided into four directional sectors that are aligned with the corresponding camera faces to enforce distance-scale alignment.

Because the Orin exposes a single PCIe Gen4 root port, a PCIe Gen3 switch-based expansion branches it into multiple M.2 endpoints. 
This configuration connects five Hailo-8 NPUs together with an NVMe SSD without requiring an external host. 
Four NPUs run SC-DepthV3, one for each cube-map face, while the fifth runs YOLO for object detection.
The VLM runs locally in a lightweight configuration and is triggered only by \texttt{walking\_impossible} events to generate a six-sentence route explanation.
All modules communicate over a lightweight ZMQ publish--subscribe bus that supports multiple message types, enabling flexible and asynchronous data exchange.
As a result, downstream stages always process fresh data, and a slow module cannot stall the pipeline.
The perception core operates independently of ROS. 
A ROS bridge exposes ROS topics to the host robot without affecting the internal execution timing.

\begin{table}[t]
\caption{Throughput and per-rail (CPU/GPU/SYS/NPU) peak power and temperature as modules are progressively enabled on the Jetson AGX Orin.}
\label{tab-thr-pow}
\centering
\setlength{\tabcolsep}{4pt}
\renewcommand{\arraystretch}{1.1}
\resizebox{\textwidth}{!}{%
\begin{tabular}{@{}
  l
  l r
  c c
  !{\hspace{0pt}\vrule width .3pt\hspace{0pt}}
  *{4}{c} 
  !{\hspace{0pt}\vrule width .3pt\hspace{0pt}} 
  *{4}{c} 
@{}}
\toprule
\multirow{2}{*}{Module} &
\multirow{2}{*}{\makecell[c]{AI Model\\(AI Task)}} &
\multirow{2}{*}{\makecell[c]{Resolution\\(W$\times$H$\times$B)}} &
\multirow{2}{*}{\makecell[c]{Thr./face\\(FPS)}} &
\multirow{2}{*}{\makecell[c]{Thr.}} &
\multicolumn{4}{c}{Power (Peak W)} &
\multicolumn{4}{c}{Thermal ($^\circ$C)} \\
\cmidrule(lr){6-9}\cmidrule(lr){10-13}
 & & & & & CPU & GPU & SYS & NPU & CPU & GPU & SYS & NPU \\
\midrule

IDLE & \NA & \NA & \NA & \NA &
0.4 W & 3.2 W & 6.1 W & \NA &
47.5 & 44.5 & 44.9 & 36.0 \\
\softhline

\makecell[l]{PCIe-m.2\\ext-card x2} &
 \NA & \NA & \NA & \NA &
\NA & \NA & 18.8 W & \NA &
\NA & \NA & 43.8 & \NA \\
\softhline

ERP-Dec & \NA &
\makecell[r]{1920$\times$960$\times$1 \\$\rightarrow$ 480$\times$480$\times$4} &
30.0 & 120.0 FPS &
0.8 W & \NA & 0.4 W & \NA &
48.5 & \NA & 46.5 & \NA \\
\softhline

\makecell[l]{3D-Depth\\} &
\makecell[l]{SC-DepthV3\\(Depth Est)} &
\makecell[r]{480$\times$480$\times$8 \\$\rightarrow$ 320$\times$256$\times$8} &
25.3 & \textsuperscript{*}101.3 FPS &
2.1 W & \NA & \NA & 8.2 W &
53.1 & \NA & \NA & 38.7 \\
\softhline

\makecell[l]{MOD\\} &
\makecell[l]{YOLOv8m\\(Object Det)} &
\makecell[r]{480$\times$480$\times$4 \\$\rightarrow$ 640$\times$640$\times$4} &
11.4 & 45.7 FPS &
0.3 W & \NA & \NA & 1.2 W &
53.5 & \NA & \NA & 40.5 \\
\softhline

\makecell[l]{TTS\\} &
\makecell[l]{MeloTTS\\(TTS)} &
\NA & \NA & \NA &
0.3 W & 1.2 W & \NA & \NA &
54.1 & 49.5 & \NA & \NA \\
\softhline

\makecell[l]{Path Desc\\} &
\makecell[l]{Qwen2.5-VL-7B\\(VLM)} &
\makecell[r]{640$\times$480$\times$1} &
\NA & 8.18 TPS\! &
0.3 W & 10.1 W & \NA & \NA &
56.8 & 53.5 & \NA & \NA \\

\midrule

Total & \NA & \NA & \NA & \NA &
4.2 W & 14.5 W & 25.3 W & 9.4 W &
\NA & \NA & \NA & \NA \\
\bottomrule

\addlinespace[1pt]
\multicolumn{13}{@{}l@{}}{\scriptsize
\textsuperscript{*} Throughput is reported for the four effective cube-map faces. 
The remaining four views are overlapping regions used for seam-aware blending.
} \\

\end{tabular}%
}
\end{table}

\subsection{Runtime, Power, and Thermal Behavior}
\label{subsec:runtime-power-thermal}

\Cref{tab-thr-pow} summarizes the end-to-end throughput, power consumption, and temperature as modules are progressively enabled on a single Jetson AGX Orin.
CPU, GPU, and system temperatures are obtained from the internal sensors. 
NPU and PCIe switch temperatures are measured using a FLIR One~\cite{flirone} under an ambient temperature of approximately \(24 \pm 1^{\circ}\mathrm{C}\).
Three observations can be made.

First, equirectangular to cube map decoding introduces only about 1.2 W of additional power while operating at 30 FPS per face. 
This efficiency is achieved using NVDEC, \texttt{nvvidconv}, and zero copy GStreamer pipelines.

Second, the Hailo-8 NPUs execute the computationally intensive workloads with low power consumption. Depth estimation requires about 8 W. 
Four cube map faces are processed in parallel, allowing each Hailo-8 to sustain more than 50 FPS. YOLOv8m increases power consumption by only about 1 W, leaving the CPU responsible mainly for data transfer and lightweight post processing. 
The event driven VLM (Qwen2.5-VL-7B~\cite{qwen2.5-vl}) operates at approximately 8 tokens/s and increases GPU power by about 10 W. 
TTS adds approximately 1 W. Since both modules are activated only when needed, the average GPU power remains low and the response latency is short.

Third, the PCIe infrastructure contributes a fixed system power overhead. 
Two PCIe Gen3 M.2 switches consume about 19 W regardless of workload. 
Even at maximum load, the total board power remains about 53 W without thermal throttling.

\boldemph{Takeaway:}
The proposed system simultaneously sustains 25.3 FPS per face for depth estimation and 11.4 FPS per face for object detection while maintaining total board power below 55 W and stable operating temperatures.

\begin{figure}[t]
  \centering
  \includegraphics[width=0.8\linewidth]{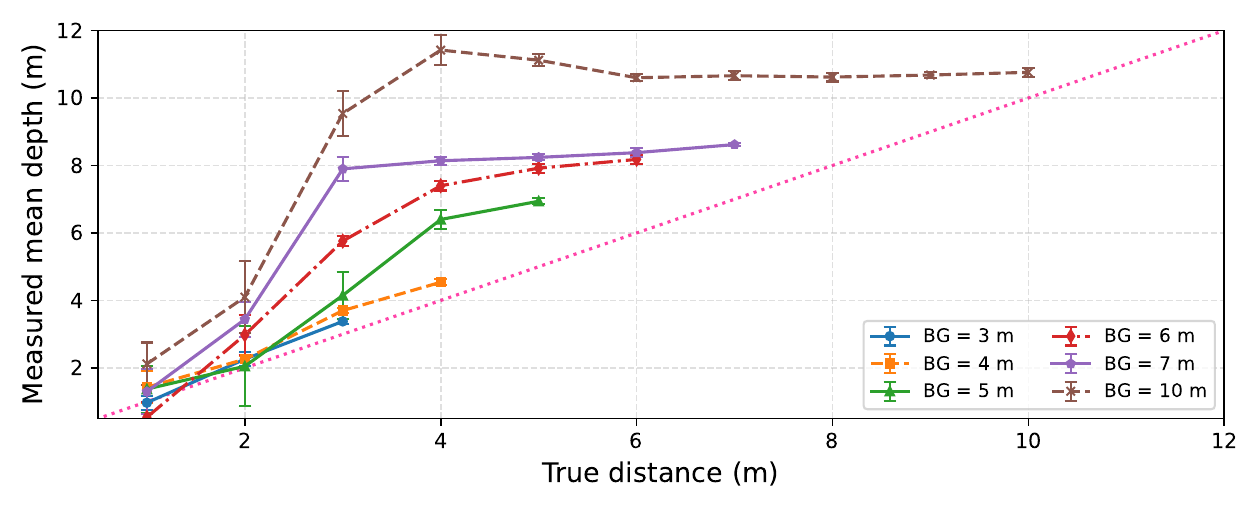}
  \caption{Measured mean depth vs.\ true distance for a person at 1--10 m under different background-wall (BG) positions (error bars: std.\ dev.; dotted: $y=x$). Near-field accuracy is high (MAE 0.34 m at 1--3 m), but mid-range estimates drift toward the wall distance.}
  \label{fig:dist-err}
\end{figure}

\subsection{Validation of Corrected Depth Estimation}
\label{depth-acc}

This experiment evaluates the reliability of the fused 360 monocular--LiDAR depth under repeatable and controlled conditions.
The background wall was placed at maximum distances of 3, 4, 5, 6, 7, and 10 meters, and a person was positioned at 1-meter intervals from the camera.
For each combination, multiple measurements were collected, and the results were summarized in terms of the mean and standard deviation.
Two consistent trends were observed.

First, the performance varies distinctly by distance range.
As shown in \cref{fig:dist-err}, the overall MAE and RMSE are 2.16 m and 2.73 m, respectively. 
Accuracy is highest in the safety-critical near range of 1--3 m (MAE 0.34 m, RMSE 0.47 m), while errors increase at 4--7 m (MAE 3.41 m, RMSE 3.95 m) and decrease again at 8--10 m (MAE 1.69 m, RMSE 1.79 m). 
This mid-range degradation indicates systematic scale drift, primarily affecting early warning rather than near-range collision avoidance.

Second, the mid-range bias is primarily caused by background dominance. 
When a distant wall occupies most azimuth columns, the mid-row affine fit is driven more strongly by the background than by a narrower foreground object, producing a characteristic ``pull-to-background'' effect. 
For example, predictions remain nearly unbiased with a 3 m wall but drift toward the wall distance as the background moves farther away.

This behavior follows from applying six angle-conditioned coefficients, estimated from mid-row LiDAR anchors, across the entire face. 
If the foreground occupies few columns or is offset from the single LiDAR scan plane, its influence on the fit is further reduced. 
The resulting bias is therefore systematic and reflects the scale limitations of the monocular depth estimate under background-dominant scenes.




\begin{figure}[t]
  \centering
  \includegraphics[width=1\linewidth]{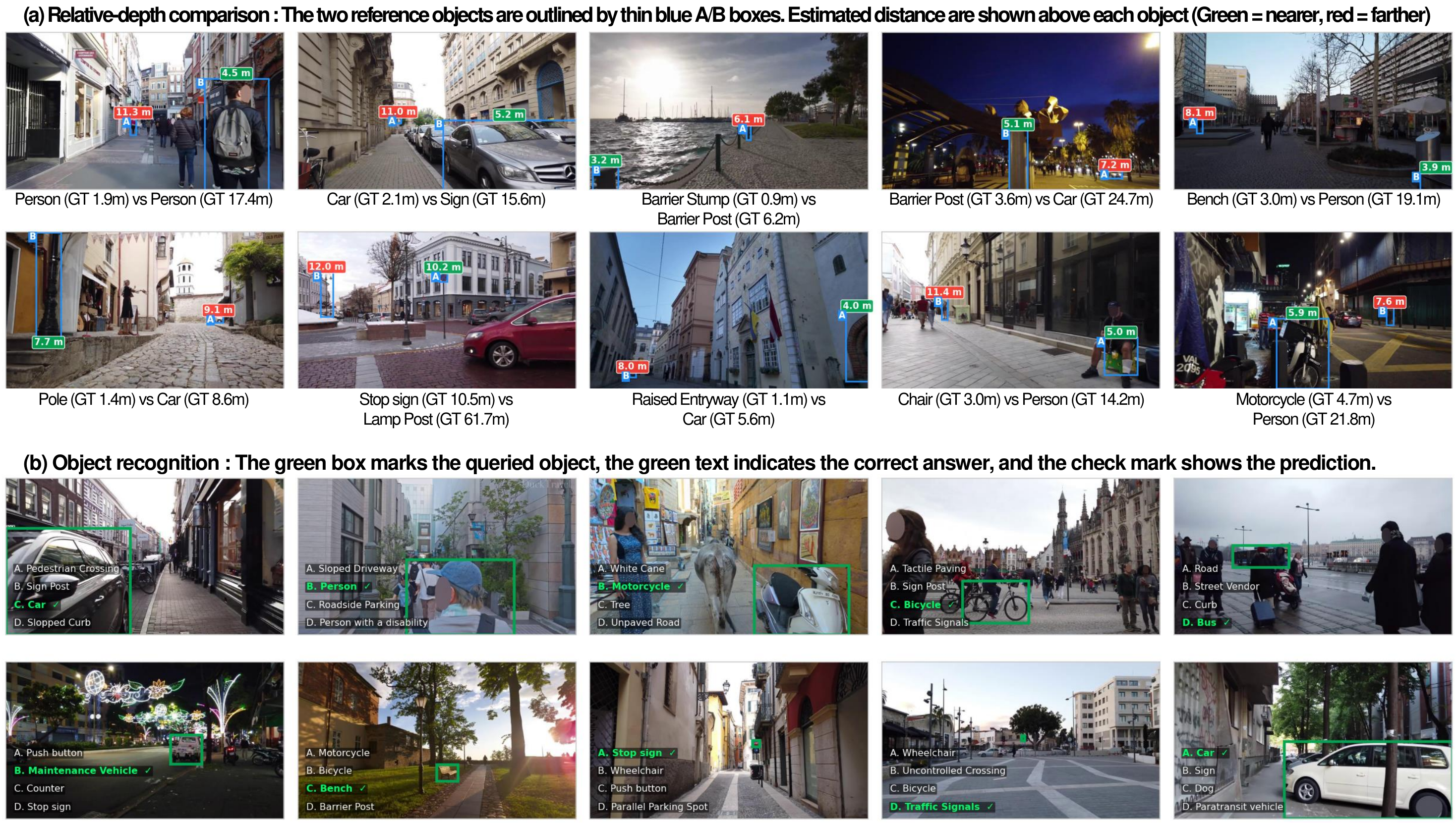}
  \caption{Qualitative GuideDogQA examples from the proposed on-device system.
(a)~Relative-depth comparison. Estimated distances are shown above the two referenced objects (green: nearer, red: farther). Ground truth distances are shown in the labels.
(b)~Object recognition. The green box marks the queried object, the green text indicates the correct answer, and the check mark denotes the predicted result.
  }
  \label{fig:guidedogqa-eval}
\end{figure}

\begin{table}[t]
\centering
\caption{GuideDogQA accuracy (\%), with a random chance level of $25.0$.
VLM baselines are reproduced from GuideDog~\cite{guidedog}. Our method (SC-DepthV3 and YOLOv8m) is evaluated as a deployed NPU implementation. 
Bold and underlined values indicate the best and second-best results, respectively. 
}
\label{tab:guidedogqa}
\small
\renewcommand{\arraystretch}{0.8}
\begin{tabular}{l c c}
\toprule
Model & \makecell{Depth\\Comparison} & \makecell{Object\\Recognition} \\
\midrule
Random Chance & 25.0 & 25.0 \\
\midrule
Cambrian-1 & 24.3 & 82.3 \\
Molmo & 28.5 & 34.0 \\
LLaVA-1.6 & 30.0 & 80.9 \\
LLaVA-OneVision & 32.4 & \textbf{87.4} \\
Qwen2.5-VL & 22.2 & \underline{85.7} \\
\quad$\hookrightarrow$ w/ finetuning & 41.5 & 83.9 \\
Gemini 2.0 Flash & 53.0 & 65.7 \\
GPT-4o & \underline{67.1} & 74.7 \\
\midrule
\textbf{Ours} & \textbf{83.8} & 63.0\textsuperscript{*} \\
\bottomrule
\multicolumn{3}{p{0.7\linewidth}}{\scriptsize
\textsuperscript{*} The COCO-80 vocabulary covers 240 of the 435 object recognition questions. Our system achieves 95.4\% accuracy on this subset.
}
\end{tabular}
\vspace{-4mm}
\end{table}

\subsection{Perception Accuracy on the GuideDogQA Benchmark}
\label{subsec:guidedogqa}

Metric depth accuracy alone does not indicate how the deployed perception modules perform in cluttered real-world environments. 
We therefore evaluate the deployed monocular depth estimator (SC-DepthV3~\cite{scdepthv3}) and object detector (YOLOv8m) on GuideDogQA~\cite{guidedog}, a real-world egocentric benchmark for blind and low-vision (BLV) guidance.

We use the two subsets corresponding to our perception modules: relative distance comparison (383 questions) and object recognition (435 questions). Random chance is $25\%$ for both tasks. All experiments are performed using the deployed models running directly on the Hailo-8 NPU.

\boldemph{(a) Relative depth comparison.}
Each question provides two ground truth bounding boxes and asks which object is closer.
A prediction is counted as correct only if both the ``closer'' and ``farther'' questions are answered correctly, resulting in a random chance of $25\%$.
We determine the answer by comparing the median predicted depth within each bounding box.
Using the ground truth boxes isolates depth estimation from object detection errors, and the resulting pairwise ordering is directly comparable with VLM-based methods.

\boldemph{(b) Object recognition.}
Each question presents four candidate object names.
We map each candidate to a COCO-80 class using a synonym table and select the class with the highest detection confidence.
Since not every answer belongs to the COCO-80 vocabulary, we report both the overall accuracy and the accuracy on the COCO-coverable subset.

\boldemph{Results.}
\Cref{tab:guidedogqa} summarizes the quantitative results, and \cref{fig:guidedogqa-eval} presents representative outputs from the deployed NPU. 
The depth estimator achieves $83.8\%$ accuracy on the relative distance comparison task, exceeding the strongest evaluated VLM, GPT-4o, which achieves $67.1\%$.
This result shows that a lightweight geometric estimator is more suitable for relative depth ordering in obstacle avoidance. 
The detector achieves $63.0\%$ overall accuracy. 
Among the 435 object recognition questions, 240 are covered by the COCO-80 vocabulary, on which our detector achieves $95.4\%$ accuracy. 
The remaining errors are primarily caused by objects outside the COCO vocabulary, such as buildings, lamp posts, and sidewalks, rather than failures of the detector itself.

\begin{figure}[t]
  \centering
  \includegraphics[width=0.95\linewidth]{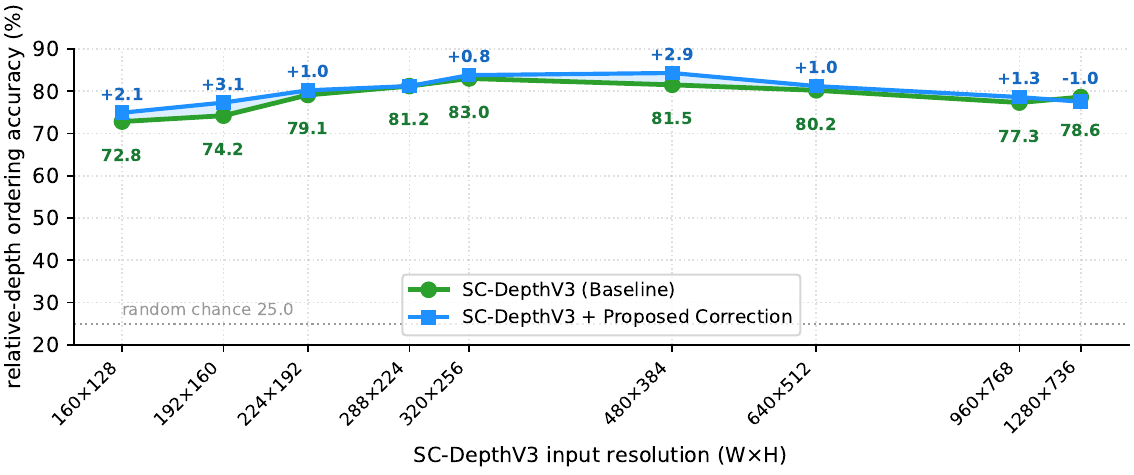}
  \caption{Effect of input resolution on relative depth ordering accuracy. 
  The proposed center-line correction improves ordering accuracy at most resolutions and achieves high accuracy at the operational resolution of $320\times256$.}
  \label{fig:res-ablation}
\end{figure}

\subsection{Depth Resolution and LiDAR-Grounded Evaluation}
\label{subsec:res-ablation}

\cref{fig:res-ablation} evaluates the effect of input resolution on depth ordering accuracy. 
SC-DepthV3 is evaluated at input resolutions ranging from $160{\times}128$ to $1280{\times}736$. 
The highest uncorrected accuracy is achieved at $320{\times}256$, which corresponds to the model's training resolution. 
Accuracy decreases at both lower ($72.8\%$) and higher ($77.3$--$78.6\%$) resolutions because these inputs fall outside the training distribution.

To evaluate the proposed center line correction across different input qualities, we apply it at every input resolution.
Since the GuideDogQA benchmark does not provide LiDAR measurements, DA-V2 is used as a high-quality reference to emulate the LiDAR center line.
We also tested Depth-Anything V3 as an alternative reference, but it resulted in lower ordering accuracy and weaker correction performance in our setting.
The proposed correction improves ordering accuracy at most resolutions, with the largest gain of $3.1$ percentage points at $192{\times}160$.
Although the corrected result at $480{\times}384$ is slightly higher than that at the deployed setting, this resolution exceeds the real-time processing budget of the on-device system.
We therefore deploy the depth estimator at $320{\times}256$, which provides the best practical balance between ordering accuracy and real-time performance.

\begin{table}[t]
\centering
\small
\caption{Relative-depth ordering accuracy and absolute depth error on 100 randomly sampled AI-Hub adverse-weather images using projected 3D LiDAR as ground truth. Bold/underlined values denote the best/second-best results.}
\label{tab:physical_gt}
\setlength{\tabcolsep}{5.0pt}
\begin{tabular}{lcc}
\toprule
Method & Acc. (\%) & MAE (m) \\
\midrule
GPT-4o & 59.8 & n/a \\
SC-DepthV3 raw & 74.3 & \underline{35.76} \\
Depth-Anything-V2 & \textbf{88.8} & 38.49 \\
Ours (DA-V2 anchor) & 72.8 & 37.29 \\
Ours (LiDAR anchor) & \underline{74.7} & \textbf{31.93} \\
\bottomrule
\end{tabular}
\vspace{-4mm}
\end{table}

\boldemph{Independent LiDAR-grounded evaluation.}
We additionally evaluate 100 randomly sampled AI-Hub~\cite{aihub} adverse-weather images against projected 3D LiDAR ground truth (565 bounding boxes and 783 relative-depth pairs).
We use unseen AI-Hub pairs instead of KITTI, which was used in DA-V2 training.
In \Cref{tab:physical_gt}, DA-V2 reaches 88.8\% ordering accuracy, supporting its use as a relative-depth proxy.
Changing the anchor from DA-V2 to physical LiDAR improves ordering accuracy from 72.8\% to 74.7\% and reduces MAE from 37.29\,m to 31.93\,m.
Compared with raw SC-DepthV3, the LiDAR-anchored result also achieves slightly higher ordering accuracy while reducing the absolute depth error.


\section{Conclusions}
We presented a lightweight and fully on-device perception system for robotic guide dogs. 
The proposed system combines a 360 camera with a 2D LiDAR using an angle conditioned affine model and seam aware blending. 


Controlled experiments verify accurate metric depth estimation in the near range. 
On the GuideDogQA benchmark, the deployed INT8 perception modules achieve 83.8\% accuracy on relative depth comparison, exceeding the strongest evaluated VLM (GPT-4o, 67.1\%). 
The object detector reaches 95.4\% accuracy on the COCO coverable subset. 
The deployed INT8 implementation also matches the accuracy of the full precision model, demonstrating that real time on-device inference can be achieved without sacrificing perception accuracy.

Several limitations remain. 
The proposed system inherits the mid range depth bias of monocular estimation, is constrained by the closed COCO vocabulary for static street infrastructure, and has not yet been validated through user studies with people who are blind or have low vision. 
Future work will focus on open vocabulary object detection within the NPU resource budget and real world field evaluations with BLV users.

\subsubsection{\ackname}
This work was supported by IITP grants funded by the Korea government (MSIT) (No.~RS-2024-00459797, RS-2025-02217404 and IITP-2026-RS-2026-25615995 under the AI Star Fellowship Support Program), a NIPA grant funded by the Korea government (MSIT) (No.~RS-2026-25621588), the Technology Innovation Program funded by MOTIE, Korea (No.~RS-2026-25620133), and an ETRI grant funded by the Korean government (No.~26CS1100).

%
%
\bibliographystyle{splncs04}
\bibliography{main}
\end{document}